\documentclass[fleqn,usenatbib]{mnras}

\usepackage{newtxtext,newtxmath}

\usepackage[T1]{fontenc}

\DeclareRobustCommand{\VAN}[3]{#2}
\let\VANthebibliography\thebibliography
\def\thebibliography{\DeclareRobustCommand{\VAN}[3]{##3}\VANthebibliography}

\usepackage{graphicx}	
\usepackage{amsmath}	

\title[Frequency-chirped laser guide stars]{Laser guide star return-flux gain from frequency chirping}

\author[J. Hellemeier et al.]{J. Hellemeier,$^{1}$\thanks{E-mail: jhelleme@phas.ubc.ca}
M. Enderlein,$^{2}$
M .Hager,$^{3}$
D. Bonaccini Calia,$^{4}$
R. L. Johnson,$^{5}$
F. Lison,$^{3}$
M. O. Byrd,$^{5}$
\newauthor
L. A. Kann,$^{5}$
M. Centrone$^{6}$
and P. Hickson$^{1}$ \\
$^{1}$University of British Columbia, Department of Physics and Astronomy, 6224 Agricultural Rd, Vancouver, BC, V6T 1Z1 \\
$^{2}$TOPTICA Projects GmbH, Lochhamer Schlag 19, 82166 Gräfelfing \\
$^{3}$TOPTICA Photonics AG, Lochhamer Schlag 19, 82166 Gräfelfing \\
$^{4}$European Southern Observatory, Karl-Schwarzschildstr. 2, 85748 Garching bei München \\
$^{5}$Starfire Optical Range, 3550 Aberdeen Ave SE, Kirtland AFB, NM, USA 87117-5776 \\
$^{6}$Istituto Nazionale di Astrofisica, Osservatorio Astronomico di Roma, Via Frascati 33, 00078 Monte Porzio Catone, RM, Italy \\}

\date{Accepted XXX. Received YYY; in original form ZZZ}

\pubyear{2015}

\begin{document}
\label{firstpage}
\pagerange{\pageref{firstpage}--\pageref{lastpage}}
\maketitle

\begin{abstract}
\textbf{Spectral hole burning reduces sodium laser guide star efficiency. Due to photon recoil, atoms that are initially resonant with the single-frequency laser get Doppler shifted out of resonance, which reduces the return flux. Frequency-chirped (also known as frequency-swept) continuous-wave lasers have the potential to mitigate the effect of spectral hole burning and even increase the laser guide star efficiency beyond the theoretical limit of a single-frequency laser. We investigate the return flux of frequency-chirped laser guide stars and its dependence on environmental and chirping parameters. On-sky measurements of a frequency-chirped, single-frequency laser guide star are performed at the Roque de los Muchachos Observatory on La Palma. In the experiment, a 35-cm telescope and a fast photon counting receiver system are employed to resolve the return flux response during laser frequency sweeps gaining insights into the population dynamics of the sodium layer.At a launched laser power of 16.5 W, we find a maximum gain in return flux of 22 per cent compared to a fixed-frequency laser. Our results suggest a strong dependence of chirping gain on power density at the mesosphere, i.e. laser power and seeing. Maximum gains are recorded at a chirping amplitude on the order of 150 MHz and a chirping rate of 0.8 MHz $\mu$s$^{-1}$, as predicted by theory. Time-resolved measurements during the chirping period confirm our understanding of the population dynamics in the sodium layer. To our knowledge these are the first measurements of return flux enhancement for laser guide stars excited by a single frequency-chirped continuous-wave laser. For higher laser powers, the effectiveness of chirping is expected to increase, which could be highly beneficial for telescopes equipped with high-power laser guide star adaptive optics systems, also for emerging space awareness applications using laser guide stars such as satellite imaging and ground-to-space optical communications.}
\end{abstract}




\section{Introduction}

Large ground-based optical telescopes suffer from wavefront distortion produced by optical turbulence in the atmosphere. To mitigate this effect, telescopes use adaptive optics (AO) systems \citep{babcock1953}. In AO systems, the wavefront is corrected by one or a combination of multiple deformable mirrors (DM). The wavefront distortion is typically sensed from a combination of natural guide stars (NGS) and laser guide stars (LGS). In principle, wavefront distortion could be sensed from NGS alone, but only a small fraction of the sky contains stars of sufficient brightness. The use of LGS increases sky coverage for AO systems \citep{ellerbroek1998}. Most state-of-the-art large telescopes and future extremely large telescopes use or will use sodium LGS, created by laser excitation of the D$_{2}$ transition in mesospheric sodium atoms.

In the Earth's atmosphere, neutral sodium is typically found between 80 and 120 km altitude. The full width at half maximum (FWHM) of the sodium concentration is approximately 10 km and the centroid is located at an altitude of approximately 90 km \citep{pfrommer2014}. Sodium is not the only element that can be used for LGS in the mesosphere. However, the product of element abundance and excitation cross-section is the highest for the sodium D$_{2}$ line yielding the highest LGS return flux for sodium LGS operating at 589 nm. Over the last two decades, dedicated high-power, continuous-wave (cw), narrow-linewidth (1-2 MHz) laser systems based on diode lasers and Raman fiber amplifiers were developed for this application. They allow for efficient optical pumping as well as re-pumping on the D$_{2b}$ transition, maximizing the LGS return flux for a given laser power. Consequently, these lasers have surpassed legacy technology in LGS brightness, but also in compactness, robustness and the potential for power scalability. 

LGS return flux is reduced by three effects \citep{holzlohner2010}. These are Larmor precession of the sodium atoms, which affects optical pumping, the change in the velocity distribution of the sodium atoms due to radiation pressure, also known as spectral hole burning, and saturation for high laser powers, which also decreases the effect of optical pumping. Earlier studies have addressed the effect of Larmor precession \citep{bustos2018a,bustos2018b} on the efficiency of optical pumping, showing that the effect can be reduced by polarisation modulation of a cw laser or by a pulsed laser. 

\begin{figure} [h]
\centering
\includegraphics[width= 9.0 cm]{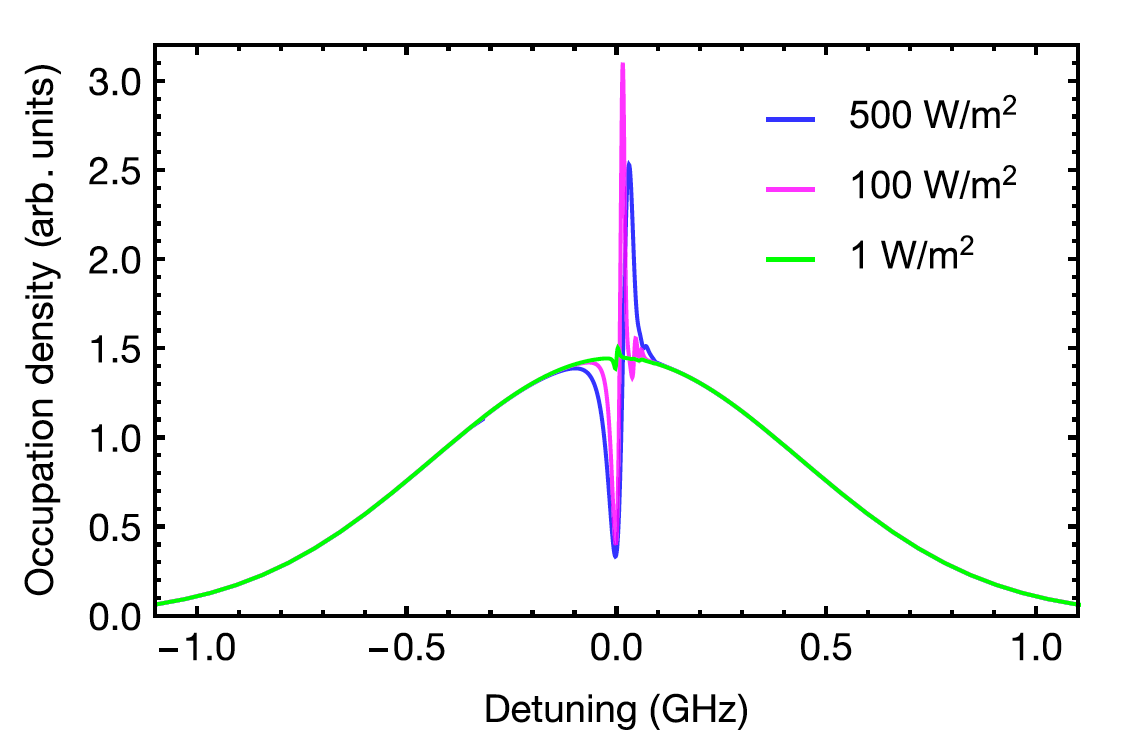}
    \caption{From \citet{Bustos2020} Computed depletion of atoms in the central velocity class of the 1-GHz FWHM mesospheric sodium line from recoil Doppler shift induced by single-frequency photons for different laser intensities. In good seeing conditions and with a 50-W cw laser, intensities of 100 Wm$^{-2}$ are typical. This depletion of sodium atoms results in a saturation of LGS brightness vs power. At 500 Wm$^{-2}$ power broadening depletes the peak in the distribution.}
    \label{fig: Spectral_Hole_Burning}
\end{figure}

Spectral hole burning denotes the effect that, under laser excitation, the velocity distribution of sodium atoms evolves into a non-equilibrium state. This is due to the net recoil resulting from repeated photon absorption and spontaneous emission cycles. Put differently, for atoms in resonance with the laser, radiation pressure increases their velocity component along the direction of the laser's Poynting vector. The corresponding Doppler shift changes the resonance frequency of the atoms as seen by the narrow-linewidth laser, depleting the number of atoms that can be repeatedly excited, as shown in Fig. \ref{fig: Spectral_Hole_Burning}. 

The velocity distribution is reset by atomic collisions on a typical timescale of 0.1 ms. To mitigate the effect on shorter timescales, the laser frequency can be dynamically shifted to follow the atomic density peak in the velocity distribution, potentially increasing the number of resonant atoms by a snow-plough effect. This technique, referred to as frequency chirping, was first suggested by \citet{jeys1991}.

Frequency chirping has also been successfully used for laser cooling of atoms, as predicted by \citet{hansch1975} and \citet{dehmelt1975} and demonstrated for trapping of Mg atoms \citep{wineland1978} and Ba$^{+}$ ions \citep{neuhauser1978}, among others. Later, the cooling of an atomic beam was demonstrated for the first time for a beam of gaseous sodium \citep{andreev1981, phillips1983, prodan1984, ertmer1985}. The method of laser cooling permitted the investigation of a hitherto unknown quantum state, the Bose-Einstein condensate \citep{anderson1995, davis1995}. For a comprehensive review of laser cooling and experiments, the reader may be referred to \citet{adams1997} or \citet{phillips1998} and references within.

Two experiments related to frequency-chirped LGS have been conducted.  \citet{kibblewhite2010} used chirped pulse trains on a 8-Watt, multiline, quasi-cw laser. In \citet{hillman2008} and \citet{Wyman2019}, two cw single frequency lasers were used, with one laser being offset from the central frequency. When both lasers were in operation and illuminating an overlapping spot in the mesosphere, the return flux did not exceed the summed fluxes of the individual lasers. As neither of these experiments used a single chirped narrow-band cw laser, the gain and effectiveness of frequency chirping for cw LGS remains uncertain.

In this paper, we report experimental results on the return flux from a LGS excited by a single-frequency chirped cw laser at 589 nm.

\section{Method}
\label{sec: method}

The median sodium column density in the mesosphere is typically about $4 \cdot 10^{13}$ m$^{-2}$, with large diurnal and seasonal variations. The temperature is about 200 K at an altitude of 90 km. The resulting Doppler linewidth of the sodium population is on the order of 1 GHz FWHM.

The sodium D$_{2}$ line arises from transitions between the $3^{2}S_{1/2}$ ground state and the $3^{2}P_{3/2}$ excited state \citep{ungar1989}. The ground state is divided into two hyperfine multiplets, which have total atomic angular momentum $F = 1$ and $F = 2$ and consist of 8 magnetic sub-states in total. The excited state consists of four multiplets having total atomic angular momenta $F' = 0,1,2,3$ and a total of 16 magnetic sub-states. The two multiplets in the ground state are separated by 1.772 GHz, whereas for the excited multiplets the separation is smaller with 15.8, 34.5 and 59.0 MHz. Transitions involving the $F = 2$ ground state constitute the D$_{2a}$ line, while the D$_{2b}$ line is generated by transitions involving the $F = 1$ ground state. For best cw sodium LGS excitation, the $m_F=2 \leftrightarrow m_F=3$ cycling transition of the D$_{2a}$ line is used, having the largest cross section and the best directional re-emission. Selective excitation of this transition is achieved by optical pumping with circularly polarized laser photons. Imperfect optical pumping, e.g. due to misalignments between laser propagation direction and the Earth's magnetic field, leads to an accumulation of atoms in the $F = 1$ ground state, where they are unavailable for re-excitation by the main laser line. This can be avoided by emitting typically $10 - 20$ per cent of the total laser power at the D$_{2b}$ frequency. This technique is called repumping and, depending on the pointing direction, it can increase the LGS return flux by a factor of 1.5-2 \citep{holzlohner2016}.

Photons carry momentum $\mathbf{p} = \hbar \mathbf{k}$, where $\hbar$ is the reduced Planck constant and $\mathbf{k}$ is the wave vector. In an absorption process the momentum of the photon is added to the initial momentum of the atom. At low saturation the atom returns to the ground state by spontaneous emission, which is isotropic on average. Hence, each scattering cycle of a D$_{2}$ line laser photon increases the momentum of the sodium atom by $1.125\cdot10^{-27}$ kg$\cdot$m s$^{-1}$ in the direction of the laser's Poynting vector, corresponding  to a spectral frequency change (Doppler shift) of 50.1 kHz. The atom is shifted out of resonance with the laser photons when the Doppler shift surpasses either the laser linewidth or, for narrow-linewidth lasers, the natural linewidth of the atom, which is $\Delta \nu_{n} = 1/(2 \pi \tau) = 10$ MHz for the sodium D$_{2}$ line with an excited-state lifetime of $\tau = 16$ ns. 

In this context, it is useful to introduce the concept of velocity classes. A velocity class consists of all atoms that can be excited by a laser whose central frequency coincides with the Doppler-shifted resonance frequency of the atoms at the centre of the velocity class. For a narrow-linewidth laser, the width of the velocity class corresponds to the natural linewidth of the atomic resonance. Atoms are shifted into the neighbouring velocity class by momentum transfer from laser photons, leading to a depopulation of the resonant velocity class and, thus, a declining return flux of scattered photons. This process is called spectral hole burning and creates a non-equilibrium atomic velocity distribution. An equilibrium distribution is restored by collisional processes, mostly with mesospheric oxygen or nitrogen atoms. The re-thermalization timescale is $\sim 0.1 ~$ms at an altitude of 90 km but is expected to vary by an order of magnitude over the extent of the sodium layer \citep{yang2021}.

Laser frequency chirping aims at remedying the spectral hole burning problem, i.e. the reduced LGS return flux caused by the depopulation of the resonant atomic velocity class. By progressively changing the laser frequency to remain resonant with the (local) maximum of the non-equilibrium atomic velocity distribution, the number of resonant atoms can be kept high, potentially even higher than in a thermal equilibrium state. This is what we refer to as snow-ploughing (or accumulating) of resonant atoms from originally different velocity classes.

The severity of spectral hole burning and, thus, the potential gains of frequency chirping, depend on the laser power density at the mesosphere, hence the combination of laser power itself and LGS spot size. In general, an LGS generated by a higher power laser will benefit more from the frequency chirping technique. Whether the potential gain can be realized depends on the chirping parameters, i.e. chirping rate $r_c$ in MHz $\mu$s$^{-1}$, chirping peak-to-peak amplitude $A_c$ in MHz, and chirping frequency $f_c = r_c / A_c$. In order to estimate a promising parameter range, we assume an atomic two-level system and an irradiance which is equal to the saturation irradiance. In this case the photon scattering rate is given by $\Delta \nu_{n}/4 = 2.5$ MHz \citep{metcalf2007}, which corresponds to one scattering event every $4/(2 \pi \Delta \nu_{n}) = 64 ~$ns. In combination with the average photon recoil of 50.1 kHz, we obtain a Doppler shift and, thus, a theoretically ideal chirping rate on the order of $r_{c,o} = 50.1 $ kHz/$ 64 $ ns $= 0.79 $ MHz $\mu$s$^{-1}$, which agrees with optimal chirp rates found in \citet{jeys1991} and \citet{Bustos2020}.

\section{Experimental setup}
\label{sec: experimental setup}

The experiment was conducted at the Roque de los Muchachos Observatory, La Palma, Spain (28$^{\circ}$45'49''N, 17$^{\circ}$53'41''W, 2396 m) using the European Southern Observatory's Wendelstein Laser Guide Star Unit (WLGSU, \citet{Bonaccini-Calia2012}) equipped with the capability for frequency chirping. The WLGSU comprises a transmitter (Tx) and a receiver (Rx), located in a sliding-roof container and a small telescope dome, respectively. The Tx consists of a 20-W-class cw 589-nm single-frequency laser and a 0.38-m refractive launch telescope mounted on an ASTELCO Systems NTM-500 direct-drive motorized mount on a fixed steel pier. 

\begin{figure}
\centering
\includegraphics[width= 9.0 cm]{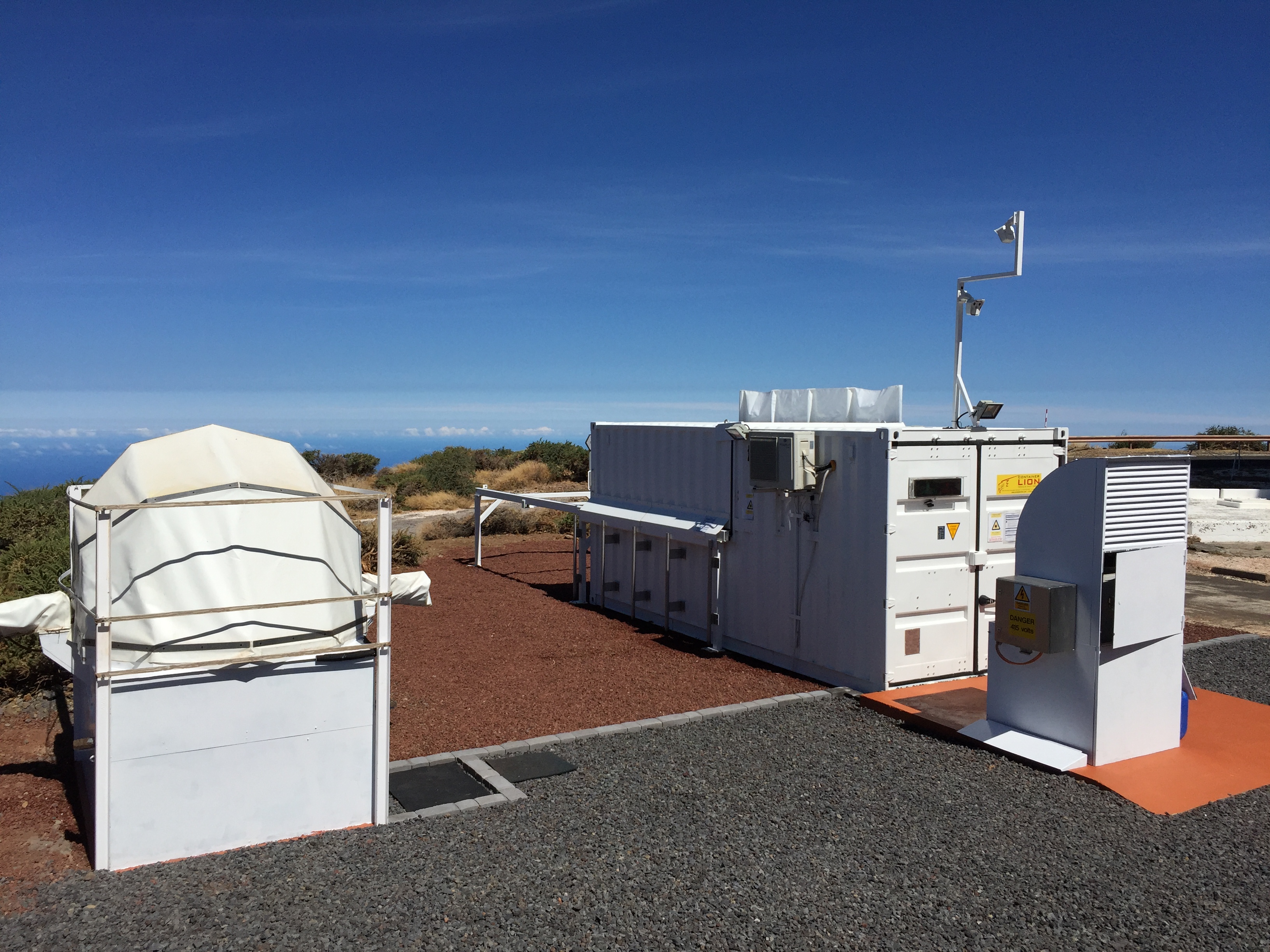}
    \caption{WLGSU configuration at Observatorio del Roque de los Muchachos next to the William Herschel Telescope, with the receiver (Rx) dome (left) and the sliding-roof container accommodating the laser beam transmitter (Tx).}
    \label{fig: WLGSU at ORM}
\end{figure}

The Rx has two independent 0.41-m and 0.36-m (C14)  receiver telescopes, binocularly configured on an ASTELCO Systems NTM-500 direct-drive mount on a steel pier. The Rx pier is at 8 m horizontal distance from the Tx pier. The Rx telescopes are equipped with SBIG photometric cameras, Johnson photometric filters, fast cameras, and one Hamamatsu H7422-40 photomultiplier operating in photon-counting mode. A reflective mirror with a central pinhole at the focal plane of the Rx Celestron C14 telescope, allows the LGS image to reach a Fabry lens followed by an Alluxa 589.16-05 ultra-narrow-band (0.5 nm)  filter, creating a telescope pupil image on the 5-mm photomultiplier. The remainder of the telescope image is reflected by the mirror to a re-imaging objective and onto a CMOS camera. The camera allows proper positioning of the LGS on the mirror pinhole to direct the light to the photon counter.  

The photomultiplier generates analog pulses for each detected photon. These are filtered and converted to TTL pulses via an Ortec 9327 discriminator. The pulse arrival times and the starting times of each chirping period were recorded with an 8-channel Roithner Lasertechnik TTM8000 time-tagging module. 

The data were acquired in multiple runs of just over a minute in length each. In the first second of each run the centroid altitude of the sodium layer was determined by modulating the laser amplitude using the WLGSU laser acousto-optic modulator (AOM) with 50 per cent duty cycle and a frequency of 10 kHz. The centroid altitude was found by cross-correlating return flux signal and modulation signal, taking into account the zenith angle. After this initial one-second sequence, laser frequency chirping was switched on an off in 15-s intervals for a total of one minute. The background flux was measured by taking a one-minute average of the flux with the telescope pointing to dark sky.

The laser of the WLGSU transmitter unit is based on a narrow-band Raman fiber amplifier operating at 1178 nm and a second harmonic generation unit \citep{calia2010}, and offers wavelength, polarization and amplitude controls. The laser is tuned to the sodium D$_{2a}$ line, at the vacuum wavelength of 589.1591 nm with a variable portion of the power shifted to the D$_{2b}$ frequency sideband (1.713 GHz separation). The AOM was used to modulate the amplitude of the emitted laser power whenever needed. The laser emits circularly polarized light in order to maximize the efficiency of optical pumping of the sodium D$_{2}$ transitions. 

To introduce a predefined chirp to the laser frequency, the laser setup is modified in the following ways: Instead of directly locking the seed laser to a wavemeter, the seed laser is frequency locked to a second, wavemeter-stabilized, laser by a fast optical phase-locked loop using a phase frequency detector. The chirp is introduced by serrodyne modulation of the frequency offset between the two lasers using the frequency modulation capability of a radio frequency function generator. The main challenge, however, is to keep the frequency doubling resonator in resonance with the chirped laser frequency. To this end up to eight adaptive feedforward controllers are used in parallel, each one acting on one of the harmonic frequencies of the sawtooth signal.

\begin{figure}
\centering
\includegraphics[width= 9.0 cm]{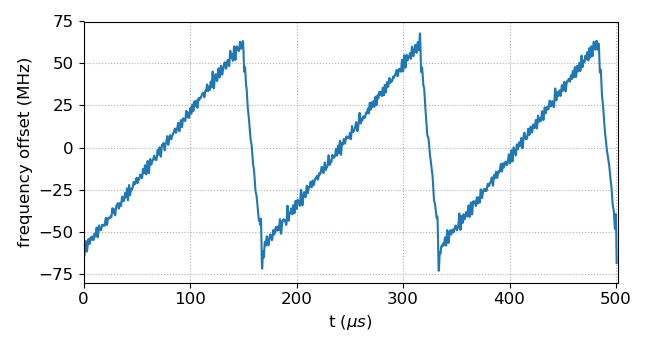}
    \caption{Frequency of the laser during chirping. In this case, the chirping frequency was 6 kHz and the chirping peak-to-peak amplitude was 120 MHz. The y-axis shows the offset from the D$_{2}$ line centre frequency. }
    \label{fig: three chirping periods}
\end{figure}

\begin{figure} [h]
\centering
\includegraphics[width= 9.0 cm]{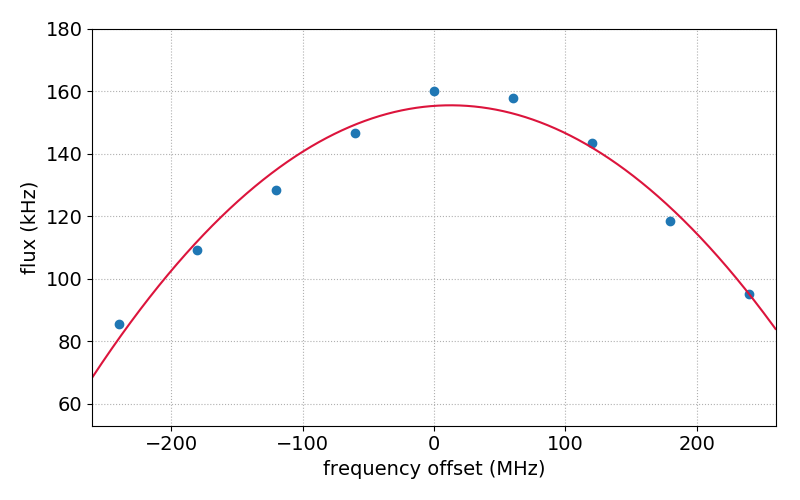}
    \caption{Measured return flux for an un-chirped laser. The true centre of the Boltzmann distribution is offset by +15 MHz from the centre frequency of the laser. }
    \label{fig: freq-scan}
\end{figure}

Due to the symmetry of the Boltzmann distribution, to first order, a linear frequency chirp that is symmetric around the sodium line center, is expected to provide the highest gains. The behaviour of the laser frequency over several chirping periods is shown in Fig. \ref{fig: three chirping periods}. At the beginning of the chirping period the laser is de-tuned below its central frequency $f_{cen}$ by half of the chirping peak-to-peak amplitude $A_{c}$. The spectral frequency is linearly increased from $f_{cen}-A_{c}/2$ to $f_{cen}+A_{c}/2$, at which point it is set back to the initial frequency and another chirping period starts. The centre frequency $f_{cen}$ is tuned to the return flux maximum, i.e. the peak of the D$_{2a}$ line, see Fig. \ref{fig: freq-scan}.  

 Ideally the duration of the frequency at the end of the chirping period is kept minimal. For our setup the time of the reset typically accounted for 10-20 per cent of the chirping period. Besides the chirping amplitude, i.e. the range of the shift in laser emission frequencies, also the chirping frequency $f_{c}$ defines the behaviour of the laser wavelength over time. The chirping frequency defines the number of chirping cycles per second and is equal to the inverse of the chirping period. The chirping rate, $r_{c} = A_{c} \cdot f_{c}$, describes the change in laser frequency per unit time. In the experiment, chirping peak-to-peak amplitudes ranging from 40 to 300 MHz and chirping frequencies ranging from 2 to 9 kHz were used, limited in both lower and higher chirping frequencies by the current design of the controller.

\section{Results}
\label{sec: results}

\begin{figure*}
\centering
\includegraphics[height=130px]{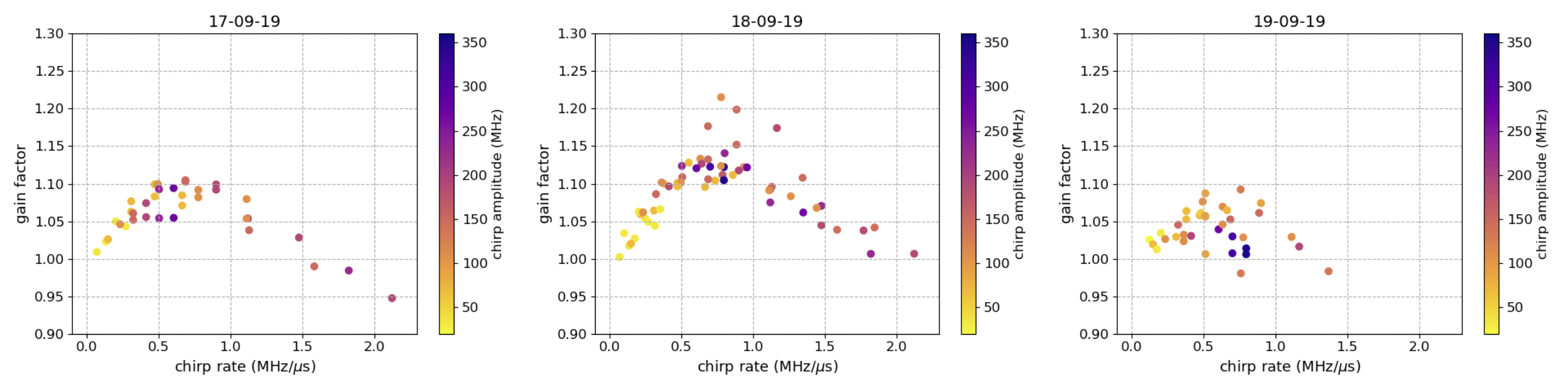}
\caption{Chirping gain vs chirping rate for the three nights of the experiment. }
\label{fig: gain_rate}
\end{figure*}

Over a period of 3 nights, 126 runs were performed, see Table \ref{tab: 1}\footnote{The seeing data in the table were provided by the Galileo National Telescope} . The runs were conducted with different chirping amplitudes $A_{c}$, chirping frequencies $f_{c}$, and chirping rates $r_{c} = A_{c} \cdot f_{c}$, in order to sweep the parameter space and find the best chirping setting in terms of maximum LGS return flux. 
The runs were therefore analyzed with regard to the gains and evolution in LGS return flux for chirped laser operation, as well as the evolution of the LGS return flux during chirping periods. 

\begin{table}[htb]
\begin{center}
\caption{Overview of observations. Median seeing values are derived from observations with a differential-image motion monitor (DIMM) located at the TNG telescope. The actual seeing at the location of our experiment is expected to be worse due to additional turbulence associated with our telescope enclosure and low-level turbulent layers not sampled by the TNG DIMM. Nevertheless, these values provide a useful indication of $relative$ seeing variations.}
\label{tab:my_label}
\begin{tabular}{c c c r}  
\label{tab: 1}
night      & runs & median seeing (") & time (UTC)~~~ \\
\noalign{\smallskip}
\hline
\noalign{\smallskip}
Sep 17 & 36 &  0.51 &  01:00+1 - 03:30+1\\
Sep 18 & 60 &  0.41 &  22:00   - 04:30+1\\
Sep 19 & 36 &  0.61 &  23:00   - 01:30+1\\
\noalign{\smallskip}
\hline
\end{tabular}
\end{center}
\end{table}

The gain in LGS return flux is of primary interest to observatories operating LGS AO facilities. It depends on properties of the atmosphere, properties of the sodium layer and chirping parameters of the laser. Studying the time-resolved behaviour of the return flux during chirping periods can reveal the efficacy of certain chirping parameters, and help further our understanding of the relevant physical processes.

\subsection{LGS return flux gain}

The LGS return fluxes for chirping and non-chirping were determined from four interlaced 15-s intervals, as described above. In each interval the first three seconds of return flux data were discarded to allow for some settling time for the laser system. The average return flux with chirping on $\langle\phi_{on}\rangle$, and chirping off $\langle\phi_{0}\rangle$, were then determined from the return flux data of the remaining 12-s intervals. Since it took a fraction of time $1-x$ of the chirping period for the laser to be reset to the start frequency, the chirping gain $g$ is not directly given by $\langle\phi_{on}\rangle/ \langle\phi_{0}\rangle - 1$. Assuming the return flux during the reset to be similar to the return flux of the un-chirped LGS, the average return flux during chirping $\langle\phi_{ch}\rangle$ is related to $\langle\phi_{on}\rangle$ and $\langle\phi_{0}\rangle$ by 
\begin{equation}
      \langle\phi_{on}\rangle = x \langle\phi_{ch}\rangle + \left(1-x\right) \langle\phi_{0}\rangle \, .  
      \label{eq: return flux} 
\end{equation}
The true gain as fraction of $\langle\phi_{ch}\rangle /\langle\phi_{0}\rangle$ is then given by\footnote{For the values of $x$, $\langle\phi_{0}\rangle$ and $\langle\phi_{on}\rangle$, which were observed in the experiment, the differences between $\langle\phi_{ch}\rangle/\langle\phi_{0}\rangle$ and $\langle\phi_{on}\rangle/\langle\phi_{0}\rangle$ are small.}
\begin{equation}
     g = \frac{ \langle\phi_{ch}\rangle}{\langle\phi_{0}\rangle} - 1  = \frac{1}{x} \left(\frac{\langle\phi_{on}\rangle}{\langle\phi_{0}\rangle}- 1\right) \, .  
      \label{eq: gain} 
\end{equation}
Our results are not corrected for a small drop in laser power which occurs when the frequency is reset from $f_{cen}+A_{c}/2$ to $f_{cen}-A_{c}/2$. Our control setup was not able to perform a fast mirror movement in an ideal way to keep the frequency doubling resonator at its resonance peak during the change in laser frequency. Thus, our LGS return flux gain values can be considered lower bounds for the actual effect of chirping. Measurements of reversed frequency chirping suggest an average reduction of the laser power by 3-8per cent.

The results for gain vs. chirping rate for three nights are shown in Fig. \ref{fig: gain_rate}. The seeing conditions were different for each night. The seeing parameter $\epsilon$ is defined as the FWHM of the image of a point source as seen through a turbulent atmosphere (with no AO compensation). For Kolmogorov turbulence, it is related to the Fried parameter $r_0$ and wavelength $\lambda$ by \citep{Fried1966} 
\begin{equation}
      \epsilon = 0.976 \frac{\lambda}{r_{0}} \, ,
      \label{eq: seeing}
\end{equation}
The up-link laser beam is also affected by atmospheric turbulence, so the seeing limits the power density that can be achieved in the mesosphere. 

Seeing values change on a second or sub-second timescale and have varied from run to run, but a general statement on the seeing conditions during the nights of observation can be made. Of the three nights, the seeing conditions were best on September 18 and worst on September 19. It is to be noted that the good seeing values reported in Table 1 were observed by the Differential Image Motion Monitor (DIMM) station of the Telescopio Nazionale Galileo. The effects of the uplink telescope aberrations and of local turbulence are not included in Table 1. 

In all nights, for low chirping rates the gain increases with increasing chirping rate, until it reaches a maximum at the optimal chirping rate, i.e. when the rate matches the corresponding recoil rate of sodium atoms. The gain falls below unity for high chirping rates as the chirping rate is much greater than the recoil rate for most atoms, which prevents efficient snow-ploughing of atoms in the velocity distribution to higher velocity classes.

\subsection{Temporal behaviour of return flux within a chirping cycle}

To study the evolution of the LGS return flux during chirping periods, return signals for many chirping periods were synchronized and averaged. Synchronization was achieved by measuring the round trip delay to the sodium profile centroid and time tagging the start of each chirping period, as described above. For averaging, each of $n$ chirping periods was divided into $m$ bins. For each of the $n\cdot m$ bins in the chirping interval the number of detected photons was determined. In this way, a bin with a number $(i-1)m + j$ with $i \in [1...n]$ and $j \in [1...m]$ contains photons that were detected in bin $j$ during chirp period $i$. By averaging over all bins with common $j$, it is possible to determine the average time-resolved return flux during the chirping period and to study its behaviour.

\begin{figure}
\centering
\includegraphics[width= 9.0 cm]{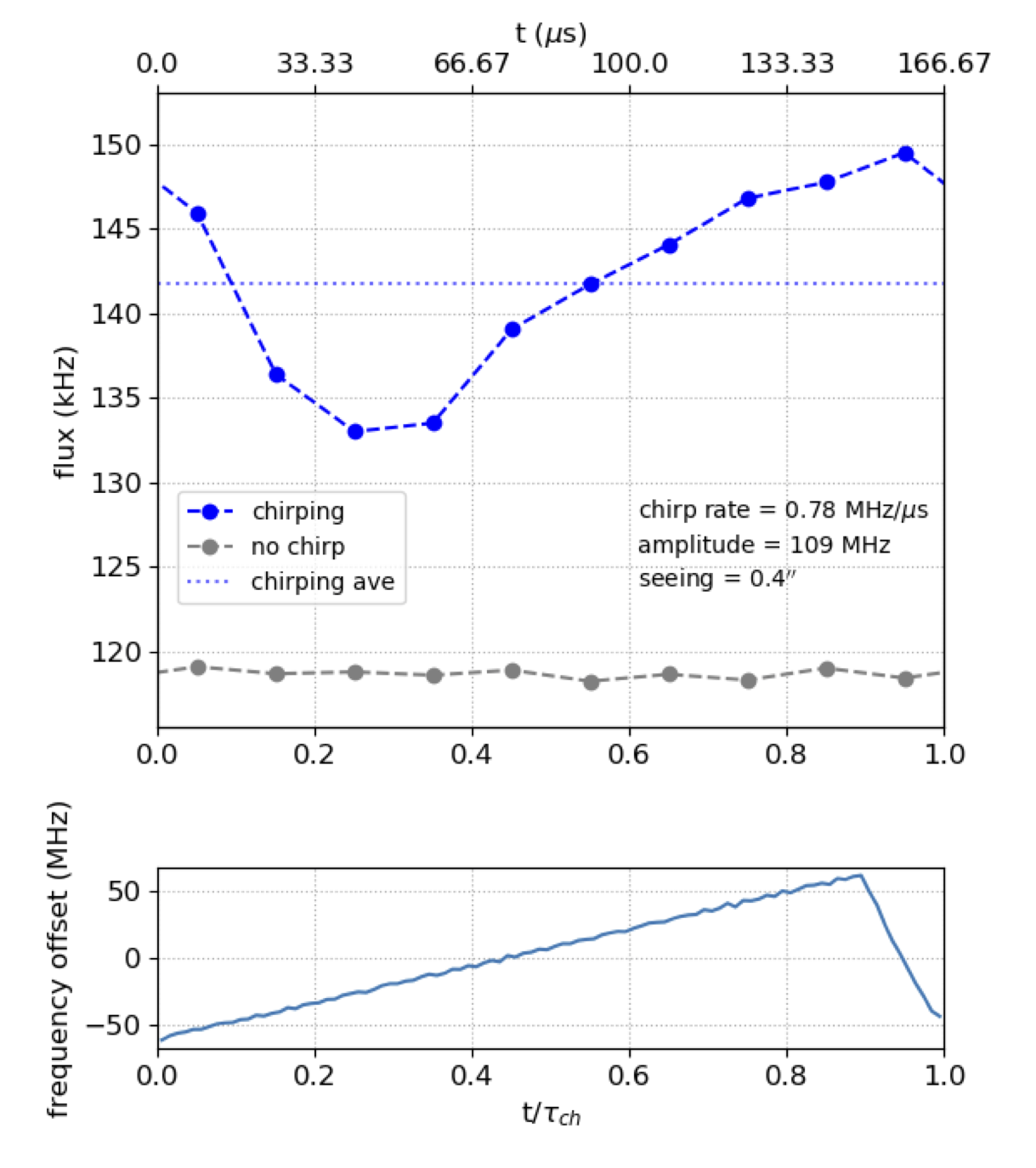}
    \caption{\textit{Top:} Time-resolved return flux during chirping periods. The data for a chirping frequency of 6 kHz and  peak-to-peak amplitude of 106 MHz are shown. The data 'chirping on' show the average return flux for  10 time bins, averaged over 72,000 chirping periods. For comparison the same averages are shown when the chirping was off.   The dotted line  'chirping ave' shows the average return flux during the chirping interval. \textit{Bottom:} Laser spectral frequency offset with respect to the central frequency over one chirping period. The shape of the curve does not follow an ideal sawtooth function. At the end of the chirp a delay of 15 per cent of the chirping period is
    required to reset the laser to the start frequency. }
    \label{fig: ramp}
\end{figure}

An example of time-resolved return flux is shown in Fig. \ref{fig: ramp}. Perhaps counterintuitively, the return flux decreases for the first 40 - 50 $\mu$s, and only then increases until the end of the chirping period. One might have expected a monotonic increase over the whole chirping period as a result of the snow-plough effect. However, this only holds for an infinitesimally thin sodium layer. In case of an extended layer, coinciding return fluxes at the detector stem from different phases of the chirping period for different altitudes. This will be discussed further in Section \ref{sec: discussion}.

\begin{figure*}[t!]
\centering
\includegraphics[height=600px, width=17.5 cm]{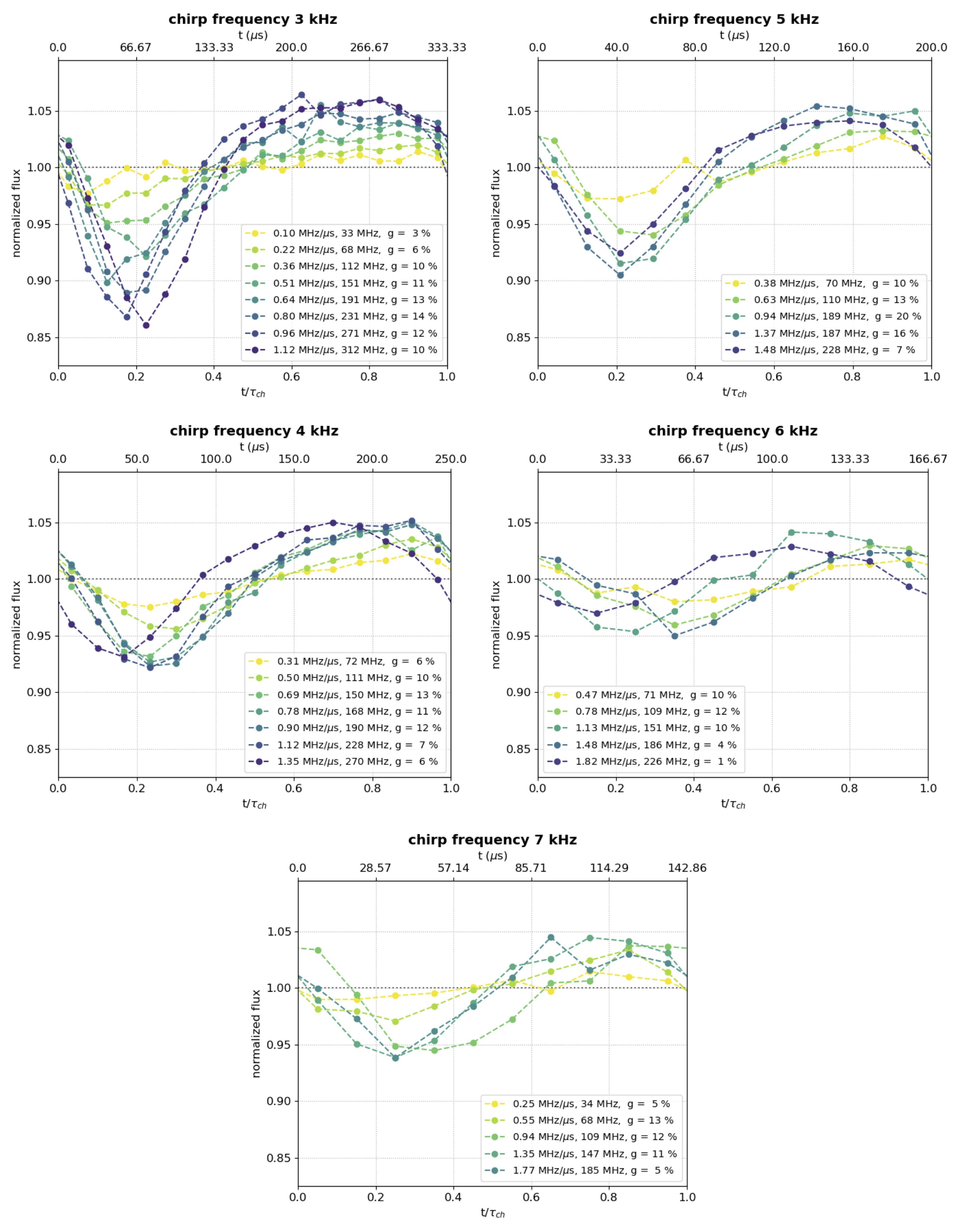}
\caption{Average return flux within a chirp for different parameters. Each plot shows the return flux for a set chirping frequency, from 3 kHz in the \textit{top left} corner to 7 kHz at the \textit{bottom}. The time bins are 16.67 $\mu$s, except for 14.82 $\mu$s at a chirping frequency of 7 kHz. The approximate elevation angle of the observations were 74$^{\circ}$, 69$^{\circ}$, 78$^{\circ}$, 70$^{\circ}$ and  77$^{\circ}$ for 3, 4, 5, 6 and 7 kHz chirping period respectively. Chirping rate, amplitude and gain $g$, for each set of parameters, are shown in the legend.}
\label{fig: intrachirp ramps}
\end{figure*}

Fig. \ref{fig: intrachirp ramps} shows the behaviour of the flux for chirping frequencies between 3 and 7 kHz with different chirping amplitudes. These data were recorded on the night that had the best seeing conditions. In order to facilitate comparison of each set of curves, fluxes are normalized by the average flux for each curve and not by the flux without chirping. The runs for a set chirping frequency were performed consecutively at similar altitudes to mitigate effects of changes in seeing conditions and in the sodium density profile and to exclude the effect of different pointing angles. The pointing for data sets having different chirping frequencies varies between $65^{\circ}$ and $80^{\circ}$ altitude.

\subsection{Effect of laser power on chirping gain}

\begin{figure}
\centering
\includegraphics[width= 9.0 cm]{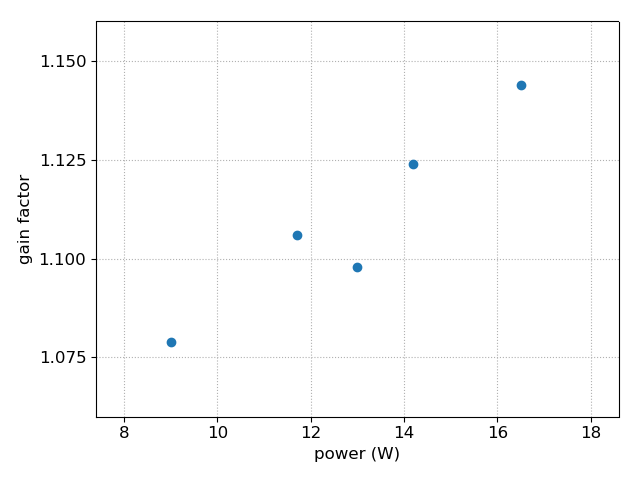}
    \caption{Chirping gain vs. laser power. The chirping frequency and chirping rate were 6 kHz and 0.78 MHz $\mu$s$^{-1}$ at an alt/az pointing of about 75$^{\circ}$/ 210$^{\circ}$. The measurements were taken consecutively over approximately 10 minutes during the night of Sep 18.}
    \label{fig: power scan}
\end{figure}

Higher laser power is expected to result in higher chirping gain. The results shown in Fig. \ref{fig: power scan} with fixed chirping frequency and amplitude and varying laser power suggest a roughly linear increase of chirping gain with laser power in the range $9 - 16.5 ~$W. This trend is confirmed by the data shown in Fig. \ref{fig: gain_rate}, which indicates increased chirping gains for better seeing conditions, i.e. higher laser irradiance at the mesosphere.

\section{Discussion}
\label{sec: discussion}

We measured a maximum chirping gain of 22 per cent for a cw laser at 16.5 W at a chirping rate of about 0.8 MHz $\mu$s$^{-1}$ on September 18 with an average seeing of 0.4". For the nights of September 17 and 19, the maximum gains are $\sim 12$ per cent and less than 10 per cent, respectively. The chirping strongly-depends on chirping rate and laser power density at the mesosphere, i.e. atmospheric seeing, launch telescope system wavefront error and laser emission power.

These results can be understood in the context of the atomic velocity distribution, shown in Fig. \ref{fig: Spectral_Hole_Burning}. At low power density, only a small fraction of the available atoms are excited, and the velocity distribution is only weakly distorted. If the power density is increased, a higher fraction of atoms is excited and spectral hole burning becomes more significant. A greater gain can then be achieved by chirping.

The power density, or equivalently, the irradiance, in the mesopshere depends both on the transmitted power and the laser spot size. The latter is affected by atmospheric seeing. Good atmospheric seeing conditions result in a smaller laser spot size and higher irradiance, leading to a greater chirping gain. 

The optimal set of chirping parameters depends on laser irradiance. In particular the optimal chirping rate is expected to be directly related to irradiance, since it depends on the rate at which the atomic population is shifted to higher velocity classes. For our results, the dependence of optimal rate on irradiance is weak, as for all three nights the optimal chirping rate is on the order of 0.7-0.8 MHz $\mu$s$^{-1}$, possibly trending towards 1 MHz $\mu$s$^{-1}$ for the night with the best seeing conditions. The values for the optimal chirp rate are close to the optimal chirp rate for a saturated two-level system of 0.79 MHz $\mu$s$^{-1}$. Once the chirp rate is increased above the optimal rate, it reaches a threshold rate at which the chirping gain drops below unity. Above this threshold value the chirping rate is too fast for an efficient snow-plowing effect. A smaller fraction of atoms are shifted to higher velocity classes, which does not compensate for the reduced atomic populations in the wings of the Boltzmann distribution. For lower irradiance, each atom is Doppler shifted less frequently, hence this threshold rate is lower. Our data suggest a strong dependence of this threshold rate on atmospheric seeing conditions, which affect the irradiance in the mesosphere. For an average DIMM seeing of 0.6" on the night of September 19, the threshold rate is between 1 and 1.5 MHz $\mu$s$^{-1}$ increasing to 2 MHz $\mu$s$^{-1}$ at 0.4" average DIMM seeing on the night of September 18.

The chirping gain should also depend on the choice of chirping amplitude. If the amplitude is reduced to zero, the chirping gain must tend to zero. If the amplitude is too large, the laser will sample velocity classes  that have a smaller atomic population.Therefore, amplitudes on the order of or larger than the width of the Boltzmann distribution of 1 GHz will result in a reduced return flux. Thus, there must be an optimal range of amplitudes. Since our setup did not allow for simultaneous seeing measurements, it is not reliable to attribute one single measurement with an outstanding gain to the chirping parameters only, more likely the result can be attributed to temporary good seeing conditions. However, two general conclusions can be drawn from the data in Fig. \ref{fig: gain_rate}. The chirping gains seem to be lower for amplitudes greater than 300 MHz and the highest chirping gains are achieved for amplitudes between 120-250 MHz, i.e. significantly below the mesospheric sodium linewidth (1 GHz). Thus, even at an optimal chirping rate it was not possible to efficiently snow-plough the complete atomic population and chirping is most efficient at the flat centre of the sodium line, where the atomic population is similar among velocity classes. This hints at the role and timescale of thermalizing collisions. In fact, the combination of 0.8 MHz $\mu$s$^{-1}$ optimal rate and 120-250 MHz optimal amplitude leads to an estimated collision time of  $0.15-0.31$ ms, which agrees with what generally has been assumed in the literature \citep[and references therein]{holzlohner2010, yang2021}. We note that sodium atoms collision rates change by about an order of magnitude over the vertical extent of the sodium layer .

The intra-chirp return flux curves in Fig. \ref{fig: intrachirp ramps} yield additional information about the population dynamics during chirping. The shapes of these curves are a combination of sodium layer effects (extension, density profile, scattering effects), chirping effects (part of the mesospheric sodium Boltzmann distribution being probed, snow-ploughing efficiency), environment effects (seeing, pointing direction) and instrumental effects (finite reset speed of the frequency chirp ramp). Without significant modelling efforts, only few qualitative conclusions can be drawn. 

For an infinitesimally thin sodium layer and optimal chirping conditions, a monotonically increasing flux would be expected starting from a value lower than average, as the laser is detuned from the centre of the Boltzmann distribution. The sodium layer extends vertically over $10-15$ km. Hence, the detected signal is a convolution of time-shifted return fluxes from different altitudes. For a sodium layer thickness of 15 km, the total time-of-flight delay between the upper and lower boundaries is 100 $\mu$s at zenith and slightly longer for elevations between 65$^{\circ}$ an 80$^{\circ}$. The curves in Fig. \ref{fig: intrachirp ramps} start roughly at the average flux. We assume higher than average flux from the end of the chirping period in the upper sodium layer and low flux from the start of the chirping period in the lower sub-layers to arrive at the detector at $t=0$.  Then the flux decreases for 40 - 60 $\mu$s$^{-1}$ as the flux of an increasing fraction of the sodium layer stems from the start of the chirping period. The decrease in return flux is strongest for high chirping rates at low chirping periods, which have the highest chirping amplitudes. At high chirping amplitudes, the laser is further de-tuned from the central frequency $f_{cen}$. Thus, fewer sodium atoms are scattering at the beginning of each chirping scan. The drop in laser power at the start/end of each chirping period is likely to amplify the effect of a decrease in flux at the start/end. Eventually the number of sublayers with increasing flux dominate, such that the total flux increases and exceeds the average flux. The curves are the steepest for chirping rates equal to or greater than the optimal chirping rate, which indicates successful snow-plowing. In the future we plan to use a second telescope, to measure the vertical emission intensity profiles and sodium density profiles at the mesosphere simultaneously, to obtain more profound knowledge of the interaction between a chirped laser and the sodium layer.

We have not studied the dependence of chirping on the pointing direction on-sky. LGS return flux depends on the angle between the laser pointing and the Earth's geomagnetic field \citep{Higbie2011} and simulations of frequency-chirped LGS show a dependence of the gain on pointing \citep{Bustos2020}. Due to the variety of environmental conditions affecting the measurements, a statistical approach involving many more measurement campaigns must be used, hence the effect has not yet been studied in detail. 
For the laser pointing at which the maximum gain of 22 per cent was achieved, simulations by \citet{Bustos2020} predict a gain of 50-55 per cent at 20 W laser power at an integrated seeing of 1". Scaling linearly with our 16 W emission we should get roughly 40 per cent gain in return flux from chirping. This discrepancy with the numerical simulations may be attributed to the different conditions along the mesosphere, and remains to be investigated.  Measurements with more parameters (e.g. integral seeing, sodium LGS vertical emission profile at the mesosphere, instantaneous laser power) monitored simultaneously, will be carried out at WLGSU.


\section{Conclusions}

We have demonstrated via on-sky observations an enhancement of LGS return flux using a chirped cw laser. An average increase in flux of up to 22 per cent was found, for a transmitted laser power of 16.5 W. High time resolution measurements have revealed how return flux increases during the chirping cycle, and how it depends on chirping parameters. Our results confirm that the chirping efficiency depends on the chirping rate, as predicted by theory and simulations. We observed the highest gain in return flux near the predicted optimal chirping rate. 

Our results suggest that the chirping efficiency depends strongly on laser irradiance at the mesosphere, hence also on atmospheric seeing conditions. For a laser power of 16.5 W, a gain greater than 10 per cent might only be observable for a seeing of 0.5 arcsec or better. However this might be an underestimate as, in our experiment, there was a small drop in laser power during the chirping cycles, which we have not accounted for in the analysis.  

The gain is found to increase linearly with laser power, at least up to 16.5 W. We therefore expect that higher-power lasers will achieve higher chirping gain. For very high irradiance, true saturation effects can set in and the response will become non-linear. Even in this case, we expect that there will be a gain for a chirped LGS. Thus, chirping could provide a significant advantage for high-power laser systems, and also improve the performance of LGS-AO systems that use uplink pre-compensation.

To match the observations with the numerical models, more parameters have to be observed or derived simultaneously and concurrently, in particular laser irradiance at the mesosphere and the vertical distribution of the mesospheric sodium emission - which we plan to do in future observing runs.  

All applications requiring high-power lasers at good observing sites may benefit from chirping, which is the only method capable of reducing the spectral hole burning effect, thus increasing the LGS return flux.

Among these applications we quote LGS-AO systems used with long pulsed lasers (tens of microseconds) at high pulse power, fast LGS-AO systems with high CW laser power, e.g. for astronomical instruments working in the visible band, and/or under harsh seeing conditions, for ground-to-space optical communication systems using LGS-AO to solve the point-ahead problem \citep{sodnik2009}, and more in general for space situational awareness LGS-AO systems applications.       

\section*{Acknowledgements}
We thank the research group of Prof. Dimitry Budker at the Helmholtz-Institute Mainz for lending us their telescope. PH acknowledges financial support from the Natural Sciences and Engineering Research Council of Canada, RGPIN-2019-04369.

\section*{Data availability}

Data available on request.


\bibliographystyle{mnras}
\bibliography{main}

\bsp	
\label{lastpage}
\end{document}